# Performance Metrics and Design Parameters for a Free-space Communication Link Based on Multiplexing of Multiple Orbital-Angular-Momentum Beams

Guodong Xie[1]*, Long Li[1], Yongxiong Ren[1], Hao Huang[1], Yan Yan[1], Nisar Ahmed[1], Zhe Zhao[1], Martin P. J. Lavery[2]
Nima Ashrafi[3], Solyman Ashrafi[3], Moshe Tur[4], Andreas F. Molisch[1], Alan E. Willner[1]
1. Department of Electrical Engineering, University of Southern California, Los Angeles, CA 90089, USA
2. School of Physics and Astronomy, University of Glasgow, Glasgow, G12 8QQ, UK
3. NxGen Partners, Dallas, TX 75219, USA
4. School of Electrical Engineering, Tel Aviv University, Ramat Aviv 69978, ISRAEL
*Email: guodongx@usc.edu

*Abstract*—We study the design parameters for an orbital angular momentum (OAM) multiplexed free-space data link. Power loss, channel crosstalk and power penalty of the link are analyzed in the case of misalignment between the transmitter and receiver (lateral displacement, receiver angular error, or transmitter pointing error). The relationship among the system power loss and link distance, transmitted beam size and receiver aperture size are discussed based on the beam divergence due to free space propagation. We also describe the trade-offs for different receiver aperture sizes and mode spacing of the transmitted OAM beams under given lateral displacements or receiver angular errors. Through simulations and some experiments, we show that (1) a system with a larger transmitted beam size and a larger receiver aperture is more tolerant to the lateral displacement but less tolerant to the receiver angular error; (2) a system with a larger mode spacing, which uses larger OAM charges, suffers more system power loss but less channel crosstalk; thus, a system with a small mode spacing shows lower system power penalty when system power loss dominates (e.g., small lateral displacement or receiver angular error) while that with a larger mode spacing shows lower power penalty when channel crosstalk dominates (e.g., larger lateral displacement or receiver angular error); (3) the effects of lateral displacement and receiver angular error are not necessarily independent; as an example of them combined, the effects of the transmitter pointing error on the system are also investigated.

## I. INTRODUCTION

Free-space communication links can potentially benefit from the simultaneous transmission of multiple spatially orthogonal beams through a single aperture pair, such that each beam carries an independent data stream and the total capacity is multiplied by the number of beams [1-6]. Orthogonality of the beams enables efficient multiplexing and demultiplexing at the transmitter and receiver, respectively.

The use of orbital angular momentum (OAM) beams as an orthogonal modal basis set for multiplexing has received recent interest. Previous experimental reports have demonstrated Terabit/s free-space data transmission using OAM multiplexing with a link distance of ~1 m [3].

With OAM, each beam has a phase front that "twists" in a helical fashion, and the beam's OAM order determines the number of $2\pi$ phase shifts across the beam [7]. Such OAM beams have ring-shape intensity distribution and phase front of $\exp(i\ell\phi)$, where $\ell$ is the topological charge and $\phi$ is azimuthal angle.

Important characteristics of each OAM beam include: (1) the intensity has a "doughnut" shape with little power in the center, and (2) the diameter of the beam grows with a larger OAM order. Moreover, the amount of phase change per unit area is greatest in the center of the beam, and phase distribution is critical for ensuring modal purity and beam orthogonality.

For a practical system, the above characteristics of the OAM beam present several important challenges when designing a free-space communication link, such as: (1) enough power and phase change of a signal need to be recovered, (2) system need to be within a durable amount of inter-modal crosstalk. An important goal that has not been adequately explored in depth is to find the systems limitations, trade-offs and design parameters for an OAM multiplexed free-space communication link [8-10].

In this paper, we explore performance metrics and design parameters for a free-space optical (FSO) communication link using OAM multiplexing. The design issues for the transmitted beam size, receiver aperture size, and mode spacing are given through the investigation of beam divergence and system power loss, channel crosstalk, and system power penalty. By analyzing power loss of the desired OAM channel due to beam divergence under a given limited-size aperture, a design consideration for the transmitted beam size is proposed. Through studying the effects of the misalignment between the transmitter and receiver (lateral displacement or receiver angular error) on OAM channel crosstalk and system power penalty, proper aperture sizes and mode spacing of the transmitted OAM beams could be selected to reduce system performance degradation. Our simulations and some experiments indicate that: (1) a system with a larger beam size and a larger receiver aperture shows a better tolerance to the lateral displacement but is less tolerant to the receiver angular error; (2) the selection of mode spacing of such a system could be based on a trade-off between signal power loss and crosstalk. For instance, a system with small mode spacing shows a lower system power penalty under a small lateral displacement or receiver angular error, while a larger mode spacing shows a lower power penalty when the lateral displacement or receiver angular error is large; (3) the effects of lateral displacement and receiver angular error are not necessarily independent. We use the transmitter pointing error as an example of the combination of lateral displacement and receiver angular error to analyze its effect on the system performance. Besides, the link design parameters for a millimeter wave (mm-wave) link using OAM multiplexing are also presented in Section IX.

In Section II, we begin with the system model. In Section III, OAM beam divergence and signal power loss within a limited size aperture are investigated. In Section IV and V, the effect of the lateral displacement, receiver angular error and transmitter pointing error on the channel crosstalk and system power penalty are analyzed, respectively. In Section VI, the system performance with the presents of both lateral displacement and receiver angular error is discussed. Section VII shows our experimental validation of the simulation model. Section VIII gives the comparison between SPP based OAM modes and LG OAM modes. Section IX describes the link design parameters for mm-wave OAM link. Section X and XI are the discussion and summary.

## II. SYSTEM MODEL

### A. Concept and simulation model

Figure 1 shows a schematic of a free-space communication link using OAM multiplexing. The multiplexed OAM beams diverge when transmitted through free space. By careful choice of the transmitted beam size, OAM mode spacing and the receiver aperture size, the system power loss, channel crosstalk and system power penalty could be reduced.

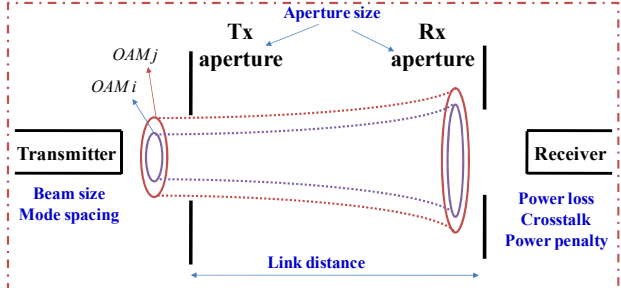

Figure 1. Concept of OAM multiplexed free-space communication link

Our simulation model of an OAM multiplexed free-space communication link is depicted in Fig. 2. Independent data streams are carried by different collimated Gaussian beams at the same wavelength, each of which is coupled from a single mode fiber to free space by a collimator. Each collimator is followed by a spiral phase plate (SPP) with a unique order to convert the Gaussian beam into a data-carrying OAM beam [11] (see Fig. 2(b)). An SPP is defined by its thickness, which varies azimuthally according to

$$h(\phi) = \phi \ell \lambda / 2\pi(n-1). \quad (1)$$

Its maximum thickness difference is $\Delta h = \ell\lambda/(n-1)$. Here, $\phi$ is the azimuthal angle varying from 0 to $2\pi$, $n$ is the refractive index of the plate material, and $\lambda$ is the wavelength of the laser beam. Different orders of OAM beams are then multiplexed to form a concentric-ring-shape and coaxially transmit through free space. The multiplexed OAM beams are numerically propagated by using the Kirchhoff-Fresnel diffraction integral [12] to the receiver aperture located at a certain propagation distance. To investigate the signal power and crosstalk effect on neighboring OAM channels, the power distribution among the different OAM modes is analyzed through the modal decomposition approach, which corresponds to the case where the received OAM beams are demultiplexed without power loss and the power of a desired OAM channel is completely collected by its receiver [13].

An experiment with a transmitted beam size of 2.2 mm over a 1-m link is carried out to partially validate our system model. In the experiment, spatial light modulators (SLMs) are used to function as SPPs at the transmitter. At the receiver, the beams are demultiplexed by another SLM loaded with an inverse spiral phase pattern of the desired mode to be detected and the resulting angularly flat phase front beam is then coupled into a single mode fiber for power measurement [14]. Assuming perfect fiber coupling, this process of OAM beam detection closely corresponds to the modal decomposition approach in our simulation model.

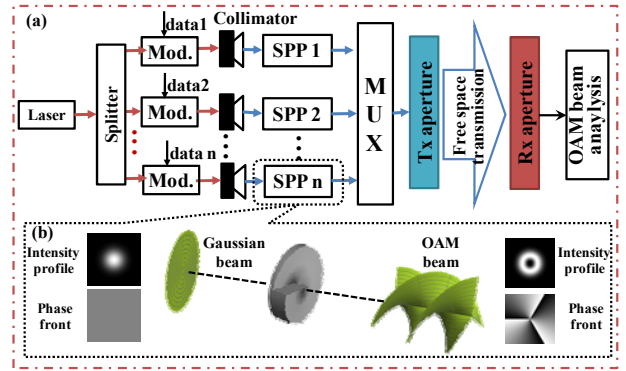

Figure 2. (a) Simulation of an OAM multiplexed data link. (b) Conversion from a Gaussian beam into an OAM+3 beam using an SPP+3 which causes helical phase shift from 0 to 6π. Tx: transmitter; Rx: receiver; SPP: spiral phase plate.

### B. Assumptions

For the convenience of analysis, the following assumptions are made:

- The wavelength of the laser source is 1550 nm. It should be noted that the specific values in the analyzed results using other wavelengths might be different. However, our fundamental approach remains valid.
- All channels have the same transmitted power.
- The collimator output at the transmitter is assumed a fundamental Gaussian beam (i.e. OAM 0) and all the OAM beams are generated from Gaussian beams with the same beam waist.
- The SPP at the transmitter is assumed to be "sufficiently large" to encompass the whole beam.
- The transmitter aperture is considered to be larger than the beam size and we assume it has no effect on the transmitted beam. Both the transmitter beam size and the receiver aperture size are parameters in the analysis.
- The insertion loss of the multiplexer is not considered, although it adds a constant insertion loss in a practical system. Besides, the insertion loss of the SPP, which is assumed to be independent of the OAM order, is also ignored.
- For calculations of spot size (beam diameter), the second moment of the intensity of an OAM or Gaussian beams, which is generally related to the beam waist, is employed, as given by the following equation:

$$D = 2\sqrt{\frac{2\int_0^{2\pi}\int_0^\infty r^2 I(r,\phi)r\mathrm{d}r\mathrm{d}\phi}{\int_0^{2\pi}\int_0^\infty I(r,\phi)r\mathrm{d}r\mathrm{d}\phi}}, \quad (2)$$

where $I(r,\phi)$ is the beam intensity profile and $(r,\phi)$ are polar coordinates [15].

- For the analysis of OAM carrying beams, we have considered Gaussian beams transformed into OAM beams by passing through SPPs (i.e. SPP based OAM beams). Most of the OAM beams used in previously reported communication links are similar to the SPP based OAM beams [3, 16, 17]. Although the OAM beams generated by passing Gaussian beams through SPPs are not exactly Laguerre-Gauss (LG) beams, such beams have similar characteristics in a communications link [18, 19]. Their difference in divergence will be further discussed in Section VIII.
- We only analyzed the case of a single-polarized system; Since there is no obvious crosstalk between different polarizations for the beam transmitted through free space, most of results could also be applied to a dual-polarization system without further modifications [3,20].

## C. Misalignment of the transmitter and receiver

In an ideal case, transmitter and receiver would be perfectly aligned, (i.e., the center of the receiver would overlap with the center of the transmitted beam, and the receiver would be perpendicular to the line connecting their centers, as shown in Fig. 3(a)). However, in a practical system, due to jitter and vibration of the transmitter/receiver platform, the transmitter and receiver may have lateral shift relative to each other (i.e., lateral displacement) or may have angular shift (i.e., receiver angular error), as depicted in Fig. 3(b) and 3(c), respectively [21]. The lateral displacement and receiver angular error might occur simultaneously. A specific example is a pointing error at the transmitter that leads to both lateral displacement and angular error at the receiver, as depicted in Fig. 3(d).

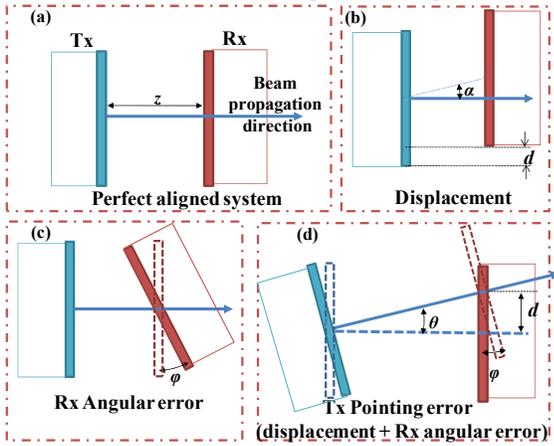

Figure 3. Alignment between the transmitter and receiver for (a) perfectly aligned system, (b) system with lateral displacement, (c) system with receiver angular error, and (d) system with transmitter pointing error. Tx: transmitter; Rx: receiver; $z$: transmission distance; $d$: lateral displacement; $\alpha$: equivalent angle for lateral displacement; $\varphi$: receiver angular error; $\theta$: pointing error.

In general, a practical link might use a tracking system to mitigate the random time-varying misalignment between the transmitter and receiver due to system vibration or long-term drift. For example, there is a commercially available tracking system with lateral resolution below 0.1 mm and angular resolution below 1 μrad [22].

TABLE I. PARAMETERS IN THE MODEL

| | |
|---|---|
| $D_t$ | Transmitted beam size (diameter) |
| $D_a$ | Receiver aperture size (diameter) |
| $z$ | Transmission distance of the link |
| $d$ | Lateral displacement |
| $\varphi$ | Receiver angular error |
| $\theta$ | Transmitter pointing error |

We analyze the performance of a free-space communication link employing OAM multiplexing for the above scenarios. The parameters discussed are listed in Tab.1.

## III. SIGNAL POWER LOSS ANALYSIS

It is generally preferred to collect as much signal power as possible at the receiver in a communications link to ensure ample signal-to-noise ratio (SNR). Since OAM beams diverge while propagating in free space and available optical elements usually have limited aperture size due to the components cost, it would be desirable to choose a proper transmitted beam size when designing an OAM multiplexed free-space communication link over a certain transmission distance. In this section, we introduce approaches to design a suitable transmitted beam size by presenting our analyses of the OAM beam divergence and power loss over different transmission distances due to limited-size apertures.

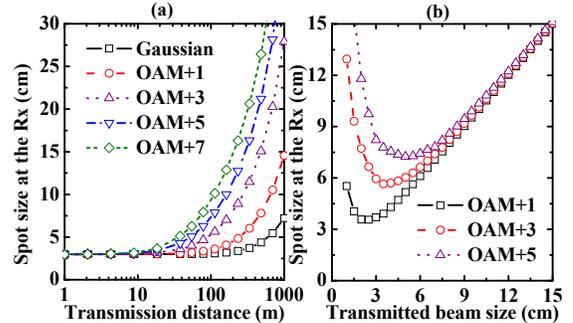

Figure 4. (a) Simulated spot size (diameter) of different orders of OAM beams as a function of transmission distance. The transmitted beam size is 3 cm. (b) Simulated spot size different order of OAM beams at the receiver as a function of the transmitted beam size for a 100-m link. The receiver aperture is not considered in this figure.

Given a fixed transmitted beam size, an OAM beam with a higher order has a larger spot size over a given distance. Figure 4(a) shows the beam divergence of different OAM beams after propagation when all transmitted beam sizes are 3 cm. The simulation results indicate that the spot sizes of these beams increase rapidly after a propagation distance of ~100 m. Figure 4(b) shows the divergence of different OAM beams when they have different transmitted beam sizes over a 100-m link. Take OAM+3 as an example: when the transmitted beam size is less than 3 cm, the spot size at the receiver increases when increasing the transmitted beam size. This is because smaller beam diffract faster. However, when the transmitted beam size is larger than 3 cm, further increasing the transmitted beam leads to larger spot size at the receiver. This is because the geometrical characteristics of the beam dominates over the diffraction. Such a trade-off needs to be considered to control the received beam size at a proper range when designing a link.

One of the effects caused by a limited-size receive aperture is signal power loss of the system, because the spot size of the diverged beam is too large to be fully captured. Figure 5 shows the power loss of OAM +3 with different transmission distances and various transmitted beam sizes. With a fixed transmitted beam size, power loss of a 1-km link is higher than that of 100-m and 10-m links due to larger beam divergence. Besides, the transmitted beam size also makes a difference. As an example, in a 100-m link, a beam with a 3-cm transmitted beam size suffers less power loss than does a beam with a 1-cm or 10-cm beam size.

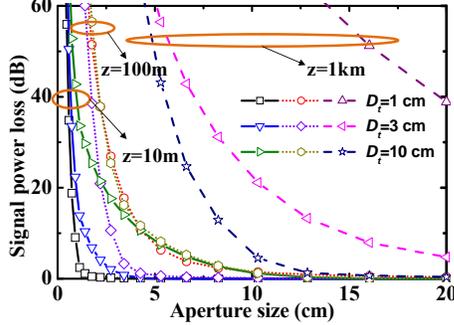

Figure 5. Simulated power loss as a function of receiver aperture size (diameter) when only OAM+3 is transmitted under perfect alignment. $D_t$: transmitted beam size; z: transmission distance.

In the following sections, we present a 100-m OAM multiplexed fre-space communication link as an example to introduce the design considerations. Here, we take 3 and 10 cm as examples for transmitted beam sizes in the link design. Our following approach could also be applied to other transmission distances and other transmitted beam sizes.

## IV. CROSSTALK AND POWER ANALYSIS

If the transmitter and receiver are perfectly aligned, the power of the transmitted OAM mode does not spread into neighboring modes, because the helical phase distribution, even within a limited-size receiver aperture, still ensures the orthogonality among different modes [23]. However, in a practical system, the lateral displacement and receiver angular error between the transmitter and receiver increases the signal power loss and also causes power leakage to the neighbors of the desired mode, resulting in channel crosstalk.

### A. Crosstalk analysis for the system with lateral displacement

First, we investigate the effect of lateral displacement on the channel crosstalk by fixing the receiver aperture size. Then, under given lateral displacements, the influence of the receiver aperture size on the system performance is studied by analyzing the power distribution among different OAM modes.

Figure 6(a) shows the power distribution among different OAM modes due to a lateral displacement between transmitter and receiver when only OAM+3 is transmitted. The transmitted beam size is 3 cm and the receiver aperture size is 10 cm. As the lateral displacement increases, the power leakage to the other modes increases while the power on OAM+3 decreases. This is because larger displacement causes larger mismatch between the received OAM beams and receiver. The power leaked to OAM+2 and OAM+4 is greater than that of OAM+1 and OAM+5 due to their smaller mode spacing with respect to OAM+3. Figure 6(b) shows much better results for the case when the transmitted beam size and receiver aperture size are 10 and 30 cm, respectively. While in both cases the Rx apertures are large enough to fully collect the corresponding transmitted beams, the larger spot size for the 10cm beam (see Fig. 4(b)) has the helical phase spread over a larger spatial scale, making the relevant mode decompositions more immune to a given lateral displacement. In a practical link design, the trade-off between the power leakage and receiver aperture size should be considered.

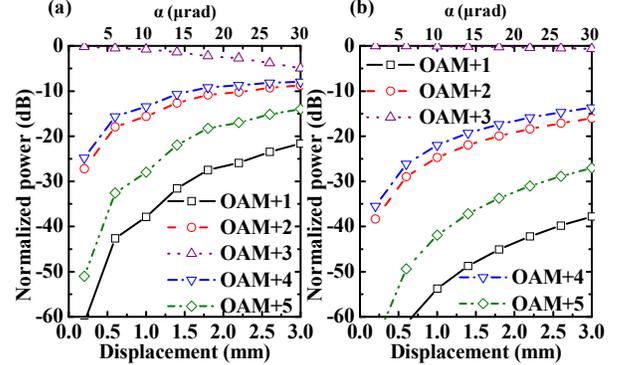

Figure 6. Simulated power distribution among different OAM modes as a function of lateral displacement over a 100-m link where only OAM+3 is transmitted. α is equivalent angle for lateral displacement. (a) The transmitted beam size is 3 cm and the receiver aperture size is 10 cm. (b) The transmitted beam size is 10 cm and the receiver aperture size is 30 cm.

Figure 7 shows the power distribution of different OAM modes under a lateral displacement of 1 mm when only OAM+3 is transmitted with a beam size of 3 cm. The results indicate that with a small receiver aperture size (less than 3 cm), the difference between power leakage to other modes and power on the desired mode is small. With a larger receiver aperture size, this power difference could be increased. However, when the receiver aperture size is large enough, the power difference increases slightly when the receiver aperture size further increases. This is because the lateral displacement could cause a mismatch of the phase profile between the received beam and the receiver, and simply increasing the receiver aperture size would not correct it.

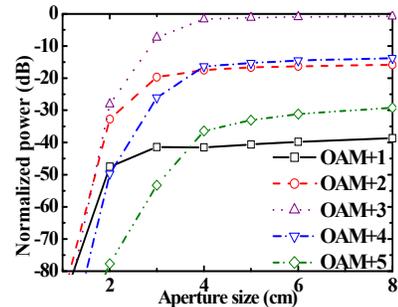

Figure 7. Simulated power distribution among different OAM modes as a function of receiver aperture size when only OAM+3 is transmitted over a 100-m link. The transmitted beam size is 3 cm and the lateral displacement is 1 mm.

### B. Crosstalk analysis for the system with angular error

Besides lateral displacement, angular errors might also occur at the receiver. In the presence of a receiver angular error of magnitude φ, the incoming phase front hitting the receiver has an additional tilt-related term and its values on the edges of the beam form $\ell\phi \pm \varphi D/2$, where $\ell$ is the topological charge and $\phi$ is azimuthal angle and $D$ is the spot size at the receiver. Clearly, these phase deviations from pure helicity are bound to introduce power leakage.

Figure 8(a) shows the power distribution among different OAM modes under different receiver angular errors when only OAM+3 is transmitted. The transmitted beam size is 3 cm and the receiver aperture size is 10 cm. With a fixed receiver aperture size, a larger receiver angular error causes a higher power leakage to the other modes. Due to the larger spot size, characterizing the case of 10-cm transmitted beam size and 30-cm receiver aperture size, poorer performance is displayed in Figure 8(b).

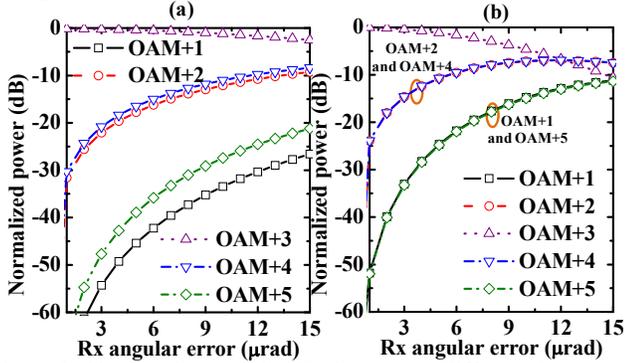

Figure 8. Simulated power distribution among different OAM modes as a function of receiver angular error over a 100-m link where only OAM+3 is transmitted. (a) The transmitted beam size is 3 cm and the receiver aperture size is 10 cm. (b) The transmitted beam size is 10 cm and the receiver aperture size is 30 cm.

### C. A specifc example of the combination of displacment and receiver angular error: transmitter pointing error

In a practical system, lateral displacement and receiver angular error might occur simultaneously, and the amounts of lateral displacement and receiver angular error might be random. We take one specific combination of lateral displacement and receiver angular error, transmitter pointing error, as an example. A transmitter pointing error $\theta$ could be considered as the combination of a lateral displacement of $d = tan(\theta) \times z$ and a receiver angular error $\varphi = \theta$, where $z$ is the link distance.

Figure 9 shows the power distribution among different OAM modes under different transmitter pointing errors when only OAM+3 is transmitted. In Fig. 9(a), the transmitted beam size is 3 cm and the receiver aperture size is 10 cm. Given a fixed transmitter pointing error or receiver angular error, the power leakage in Fig. 9(a) is higher than that in Fig. 8(a) because a transmitter pointing error includes a lateral displacement in addition to the receiver angular error. When the transmitted beam size is 10 cm and the receiver aperture size is 30 cm, as shown in Fig. 9(b), the power distribution is similar to that in Fig. 8(a). One reason is that when the transmitted beam size is large, the effects of the receiver angular error dominate over the effects of the lateral displacement.

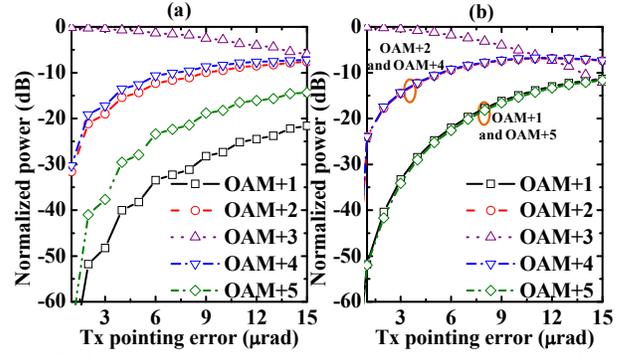

Figure 9. Simulated power distribution among different OAM modes as a function of transmitter pointing error over a 100-m link where only OAM+3 is transmitted. (a) The transmitted beam size is 3 cm and the receiver aperture size is 10 cm. (b) The transmitted beam size is 10 cm and the receiver aperture size is 30 cm.

As discussed in this section, one of the approaches for link design is that a larger transmitted beam size could help to reduce the power leakage due to lateral displacement, resulting, however, in more power leakage due to receiver angular error.

## V. POWER PENALTY ANALYSIS

A consequence of the signal power loss and channel crosstalk analyzed in the previous sections is the increase in system power penalty, which is the signal to noise ratio (SNR) difference need to achieve a certain bit error rate (BER) by an OAM channel and ideal channel. It is used as metric to evaluate the system performance degradation. Signal power loss resulting from limited-size receiver aperture and channel crosstalk due to lateral displacement or receiver angular error degrades the signal-to-interference-plus-noise ratio (SINR) of each channel, thus affecting the BER or power penalty performance. We simulated a four-channel OAM free-space communication link each channel transmitting a 16-QAM signal. Noting that other modulation formats could also be used, the results of a specific value might be different but the trends of the curves should be similar. An approach for the design the receiver aperture size and mode spacing in such a system is then introduced based on the power penalty analysis.

With the background noise assumed to follow the Gaussian model, the error probability of a 16-QAM signal is [24]:

$$P_{e,16-QAM} = 3Q\left(\sqrt{\frac{4}{5}\frac{E_{avg}}{N_0}}\right)\left[1 - \frac{3}{4}Q\left(\sqrt{\frac{4}{5}\frac{E_{avg}}{N_0}}\right)\right] \quad (3)$$

where $E_{avg}/N_0$ is the average signal-to-noise ratio (SNR) per bit. $E_{avg}$ is the average signal power per bit and $N_0$ is the power density of a Gaussian white noise. Equation (3) shows that a minimum transmitted power $P_{tx}$ is required to achieve a certain BER, given a Gaussian background noise $N_0$. In our simulation, two interleaved extended BCH(1020,988) code is considered for forward error correction (FEC); such a code results in a payload length of (522240, 489472) and an overhead of ~7%. The system need a raw BER of $3.8 \times 10^{-3}$ to achieve block error rate of $10^{-12}$ [25].

With the assumption that all channels have the same transmitted power and that channel crosstalk interferes with the signal in a similar way to noise at our BER threshold of $3.8 \times 10^{-3}$ [26], the required transmitted power $P_{tx,m}$ for channel $m$ in a multiplexed system with signal power loss and channel crosstalk could be expressed as:

$$P_{\text{tx,m}} = P_{\text{tx}} \left( \alpha - \beta \cdot \frac{P_{\text{tx}}}{N_0} \right)^{-1}, \qquad (4)$$

where $\alpha$ and $\beta$ are normalized signal power and crosstalk at the receiver, respectively. The power penalty is defined as:

$$P_{\text{penalty}} = 10 \cdot \log_{10}\left(\frac{P_{\text{tx,m}}}{P_{\text{tx}}}\right) \text{ dB} \qquad (5)$$

To explore the influence of limited-size receiver aperture and lateral displacement on power penalty, four channels are simulated in a 100-m OAM multiplexed free-space communication link. Power penalties for all four channels might be different due to different OAM orders have different spots sizes. To ensure that every channel works, the largest power penalty among all channels is defined as the system power penalty.

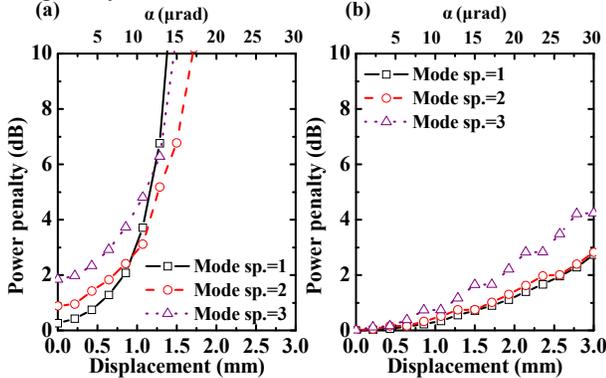

Figure 10. Simulated system power penalty as a function of lateral displacement when different sets of OAM beams are transmitted over a 100-m link. Mode spacing=1: OAM+1, +2, +3, and +4 transmitted. Mode spacing=2: OAM+1, +3, +5, and +7 transmitted. Mode spacing=3: OAM+1, +4, +7, and +10 transmitted. $\alpha$ is equivalent angle for lateral displacement. sp.: spacing. (a) The transmitted beam size is 3 cm and receiver aperture size is 10 cm. (b) The transmitted beam size is 10 cm and receiver aperture size is 30 cm.

We simulate various sets of OAM beams to analyze system power penalty with different mode spacings. Fig. 10(a) shows when the transmitted beams size is 3 cm. When the lateral displacement is larger than 0.75 mm, the system with mode spacing of two (OAM +1, +3, +5, and +7 transmitted) shows a lower power penalty than with mode spacing of one (OAM +1, +2, +3, and +4 transmitted). This is because the channel crosstalk between adjacent OAM modes is higher than between OAM modes of a spacing of two. When the lateral displacement is less than 0.75 mm, the system with mode spacing of one shows less power penalty than with mode spacing of two. This is because the system with mode spacing of two has larger power loss due to the larger beam size at the receiver. Since the system with mode spacing of three has even larger beam divergence, its power penalty is higher. Fig. 10 (b) shows the case when the transmitted beam size is 10 cm and the receiver aperture size is 30 cm. A comparison to the results in Fig. 10 (a) shows that a larger transmitted beam size could help reduce the system power penalty caused by lateral displacement.

Similarly, the influence of the receiver angular error on the system power penalty is also explored (see Fig. 11). In Fig. 11(a), the transmitted beam size is 3 cm and the receiver aperture size is fixed to 10 cm. Different sets of four OAM beams are transmitted over a 100-m link. The mode spacing of two has better performance than mode spacing of one when the receiver angular error is larger than 6 µrad. In Fig. 11(b), where the transmitted beam size is 10 cm and receiver aperture size is fixed to 30 cm, the power penalty is slightly larger than that in Fig. 11 (a). Figure 12 shows the system power penalty when there is transmitter pointing error. Figure 12(a) shows a higher power penalty than does Fig. 11(a) because the transmitter pointing error contains extra lateral displacement besides the receiver angular error. In addition, Fig. 12(b) shows a trend similar to Fig. 11(b) because when the transmitted beam size is large, the power penalty is mostly caused by the receiver angular error as compared to the lateral displacement effects.

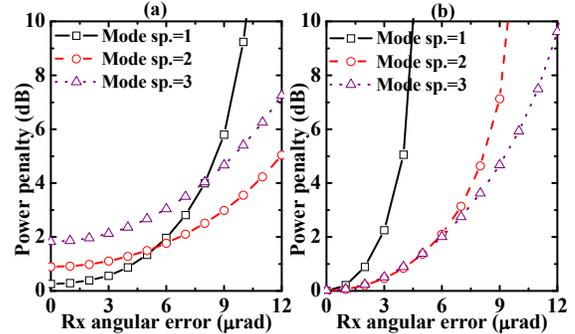

Figure 11. Simulated system power penalty as a function of receiver angular error when different sets of OAM beams are transmitted in a 100-m link. (a) The transmitted beam size is 3 cm and receiver aperture size is 10 cm. (b) The transmitted beam size is 10 cm and receiver aperture size is 30 cm. sp.: spacing.

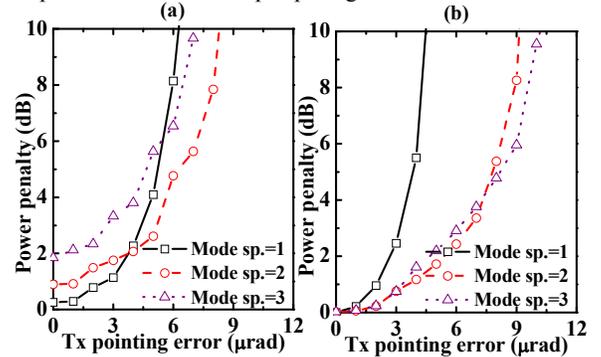

Figure 12. Simulated system power penalty as a function of transmitter pointing error when different sets of OAM beams are transmitted in a 100-m link. (a) The transmitted beam size is 3 cm and receiver aperture size is 10 cm. (b) The transmitted beam size is 10 cm and receiver aperture size is 30 cm. sp.: spacing.

The power penalty analysis indicate some selections rules for mode spacing: (a) a larger transmitted beam size and receiver aperture could increase the system tolerance to lateral displacement but decrease its tolerance to receiver angular error; (b) systems with larger mode spacing has higher order OAM beams, which leads to a higher signal power loss due to beam divergence; however, it also suffers less channel crosstalk. As a trade-off between signal power loss and crosstalk, a system with a small mode spacing shows a lower system power penalty under a small lateral displacement or receiver angular error, while a larger mode spacing shows lower power penalty when the lateral displacement or receiver angular error is large.

## VI. SYSTEM PERFORMANCE IN THE PRESNETS OF BOTH LATERAL DISPLACEMENT AND RECEIVER ANGULAR ERROR

In a practical system, the lateral displacement and receiver angular error might occur simultaneously. It is shown in the previous section that when the transmitted beam size and receiver aperture size are larger, the system exhibits greater tolerance to the lateral displacement but lower tolerance to the angular error. Given certain lateral displacements and receiver angular errors, how to select the transmitted beam size and receiver aperture size to reduce the total power penalty would be an interesting question. We fix the mode spacing to two and transmit OAM+1, +3, +5 and +7, simultaneously. Different transmitted beam sizes 5, 6, 8, 10, 15, 20, and 30 cm with corresponding receiver aperture size of 15, 18, 24, 30, 45, 60, and 90 cm are considered.

Figure 13 shows the system power penalty for different transmitted beam sizes considering lateral displacement or receiver angular error. When the lateral displacement is 3 mm, a system with the transmitted beam size of 10 cm suffers ~6 dB less power penalty than that with a transmitted beam size of 6 cm (see Fig. 13(a)). However, the former suffers 3 dB more power penalty than the latter when the receiver angular error is 10 μrad (see Fig. 13(b)). There exists a trade-off between the effects of the lateral displacement and receiver angular error. For the parameters design of a practical system, one might need to select a proper beam size to reduce the system performance degradation considering this trade-off.

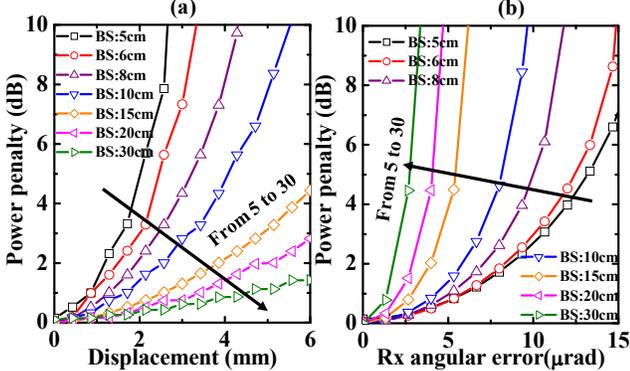

Figure 13. Simulated system power penalty as a function of (a) lateral displacement, (b) receiver angular error when the mode spacing is 2 and transmission distance is 100 meter. BS: Transmitted beam size. The receiver aperture size is three times the size of the transmitted beam size.

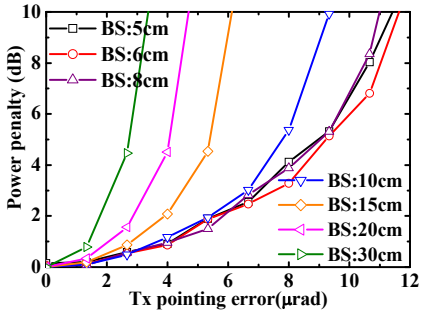

Figure 14. Simulated system power penalty as a function of transmitter pointing error when the mode spacing is 2 and transmission distance is 100 m. BS: Transmitted beam size. The receiver aperture size is three time the size of the transmitted beam size.

We also analyzed the effects of the transmitter pointing error on the system power penalty. Figure 14 indicates that a system with small transmitted beam sizes (5, 6 and 8 cm) shows similar system power penalty because neither the lateral displacement nor the receiver angular error dominates and when their effects are added to each other, the total effects for different transmitted beam size are near. We see that the power penalty shows a similar trend to that caused by only the receiver angular error at a larger transmitted beam size, because the effects of the receiver angular error dominates over the lateral displacement.

## VII. EXPERIMENTAL VALADITION OF THE MODEL

As a partial validation of our link model, an experiment without lateral displacement between the transmitter and receiver is first introduced. Figure 15(a) shows that the experimental results of power loss of different OAM modes due to limited-size receiver aperture are in good agreement with the simulation results.

Another validation of the simulation model considering a lateral displacement is shown in Fig. 15(b). Over the 1-m link and a transmitted beam size of 2.2 mm, OAM+3 is transmitted with a lateral displacement of 0.2 mm. Measured and simulated power distribution with different receiver apertures show similar trends.

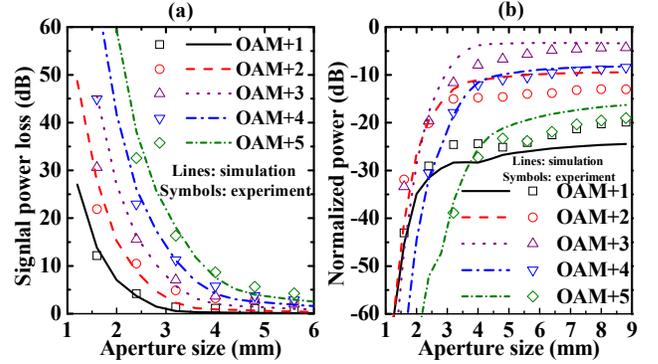

Figure 15. (a) Comparison between experimental and simulated power loss of different OAM beams as a function of receiver aperture size. The transmitter and receiver are considered perfectly aligned. (b) Comparison between experimental and simulated power distribution among different OAM modes as a function of receiver aperture size with a lateral displacement of 0.2 mm. In both figures, only OAM+3 is transmitted with the transmitted beam size of 2.2 mm over 1-m link. Lines and symbols are simulation and experiment results, respectively.

## VIII. COMPARISON OF SPP BASED OAM MODES AND LAGUERRE-GAUSS OAM BEAM

There are other modal sets which posses OAM for multiplexing. One such example is LG modes as mentioned in section II. Both an SPP based OAM beam and an LG OAM beam have similar characteristics, including helical phase front structure and doughnut intensity shape. However, a slight difference lies in their radial intensity profiles. To explore the performance differences of systems using SPP based OAM beams and LG OAM beams, Figure 16 shows the comparisons of power loss as well as power distribution over a 100-m link due to the limited-size receiver aperture and the later displacement, respectively. Both beams show similar trends, indicating their similar properties for a communication system.

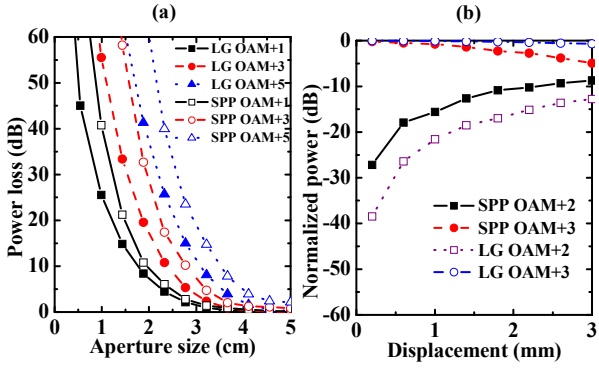

Figure 16. (a) Comparison between the power loss of SPP generated OAM beams and pure Laguerre-Gaussian beam as a function of receiver aperture size. (b) Comparison between received power of OAM+2 and OAM+3 by SPP generated OAM beams and pure Laguerre-Gaussian beam as a function of lateral displacement when only OAM+3 is transmitted. The transmitted beam size of all beams is 3 cm and link distance is 100 meter.

IX. DESIGN COSIDERATION FOR MM-WAVE SYSTEM

The above analyses in the previous sections focused on the parameter design of an OAM multiplexed free-space communication link in the optical region at 1550 nm. Recently, the use of OAM multiplexing in mm-wave communication systems for increasing system capacity has also aroused wide interests [4, 5, 28, 29]. Here, we select a mm-wave link at 90 GHz as an example to explore the link design approaches. Figure 17(a) shows the power loss as a function of the receiver aperture size under a link distance of 100 m and the transmitted beam size of 100 cm. Figure 17(b,c) shows the received power distribution (leading to channel crosstalk) among OAM channels (OAM+1, +2, +3, +4 and +5) with various lateral displacements and receiver angular error when only OAM+3 is transmitted. Figure 18(d,e,f) shows the system power penalty due to lateral displacement, receiver angular error and transmitter pointing error for different mode spacing. The trends of the curves for power loss, received power distribution and power penalty agree well with that of the optical system. However, the following points also need to be considered for a link design,

- In general, mm-wave OAM beams diverge faster than optical beams with the same order and the same transmitted beam size due to the diffraction limit. Therefore, mm-wave systems might need larger transmitted beam sizes and larger receiver apertures. However, mm-wave OAM beams could be less sensitive to lateral displacements and angular errors than optical beams due to their longer wavelengths.
- A mm-wave OAM system can use other carrier frequency, e.g., 28 GHz. As mm-waves at lower frequencies (e.g. 28 GHz) diverge more than waves at higher frequencies (e.g. 90 GHz), special considerations of beam sizes or aperture sizes might be required.
- Our results are focused on a 100-m link. It is expected that an OAM mm-wave system suffers less beam divergence for a shorter distance link (e.g. 10 meter).

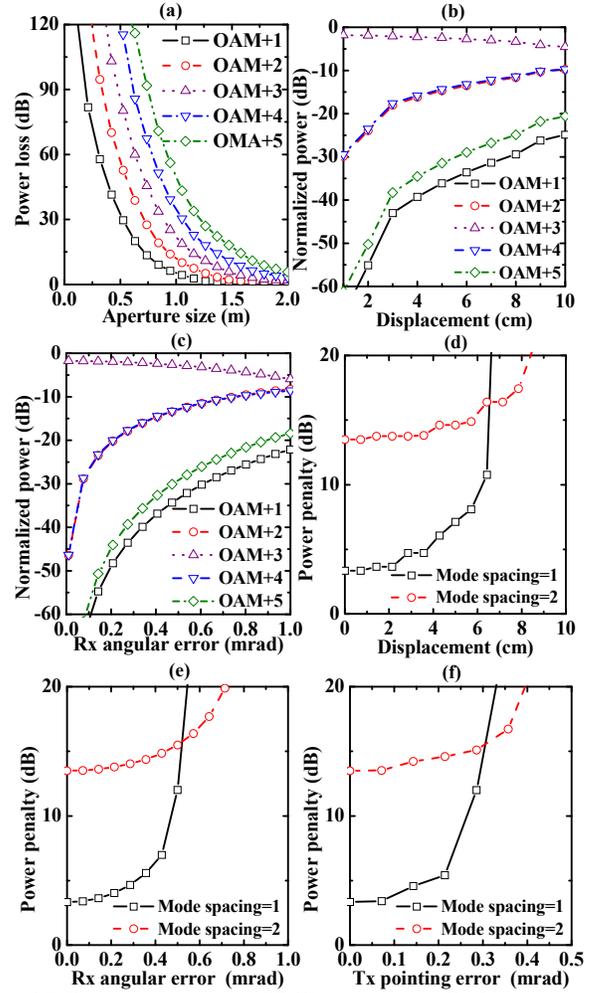

Figure 17. Simulated results of 90 GHz mm-wave OAM beams over a distance of 100 m with a transmitted beam size of 1 m. (a) Power loss as a function of receiver aperture size when the system is perfectly aligned; (b) Power distribution as a function of lateral displacement with a receiver aperture size of 2 m; (c) Power distribution as a function of receiver angular error receiver with a receiver aperture size of 2 m; (d,e,f) System power penalty as a function of lateral displacement, receiver angular error and transmitter pointing error for different mode spacing under the receiver aperture size of 2 m.

X. ADDITIONAL CONSIDERATIONS

The following points are worth mentioning:
- We only consider the use of OAM beams with plus charges for data transmission. Our design approach could be similarly applied to the system in which OAM beams with both plus and minus charges are used for multiplexing [3, 6, 20].
- Atmospheric turbulence might result in beam distortions [30-32] in a free-space link. Of all the effects caused by turbulence, beam wandering and arrival angle fluctuation effects could be considered as the lateral displacement and receiver angular error discussed in our design approach.
- Digital signal processing algorithms, such as multiple-input-multiple-output (MIMO) equalization that can be used for channel crosstalk mitigation [33-35] are not employed in our simulations. The effects

- of the lateral displacement and receiver angular error could be reduced if using such algorithms.
- In a practical case, lateral displacement and receiver angular error are generally time-varying random processes. Our approach could help provide the analysis of upper and lower bound of system performance, given a specific error dynamic range of each process.

## XI. SUMMARY

We explored performance metrics and design parameters for OAM multiplexed free-space optics as well as mm-wave communication links. The link distance, transmitted beam size, transmitter and receiver aperture sizes and OAM mode spacing were studied holistically. By analyzing the system power loss, channel crosstalk and system power penalty, a proper transmitted beam size, receiver aperture size, and OAM mode spacing could be selected for the system to handle lateral displacement, receiver angular error or transmitter pointing error between the transmitter and receiver.

## ACKNOWLEDGMENT

We thank Robert Boyd, Nivedita Chandrasekaran, Ivan Djordjevic, Prem Kumar, Mark Neifeld, Miles Padgett, Jeffrey Shapiro, and Tommy Willis for the fruitful discussions. We acknowledge the generous support of DARPA under the InPho program, Intel Labs University Research Office and NxGen Partners (solyman.ashrafi@nxgenpartners.com).